\documentclass[11pt]{article}

\usepackage{acl}

\usepackage{times}
\usepackage{latexsym}
\usepackage[T1]{fontenc}
\usepackage[utf8]{inputenc}
\usepackage{microtype}
\usepackage{inconsolata}

\usepackage{amsmath}
\usepackage{amsfonts}
\usepackage{amssymb}
\usepackage{nicefrac}
\usepackage{graphicx}
\usepackage{booktabs}
\usepackage{multirow}
\usepackage{subcaption}
\usepackage{xspace}
\usepackage{xcolor}
\usepackage{url}
\usepackage{tikz}
\usepackage{algorithm}
\usepackage{algpseudocode}

\graphicspath{{figs/}}

\newcommand{\ours}{AgentProv\xspace}
\newcommand{\oursfull}{Agentic Provenance\xspace}

\newcommand{\CR}[1]{\textcolor{black}{#1}}

\newcommand{\R}{\mathbb{R}}
\newcommand{\E}{\mathbb{E}}
\newcommand{\indicator}{\mathbf{1}}
\newcommand{\MMD}{\mathrm{MMD}}
\newcommand{\hMMD}{\widehat{\mathrm{MMD}}}

\newcommand{\policy}{\pi}
\newcommand{\refpolicy}{\policy_{\text{ref}}}
\newcommand{\suspolicy}{\policy_{\text{sus}}}
\newcommand{\fingerprint}{\varphi}

\title{\ours: Auditing Agentic LLM API Providers via\\Tool-use Policy Probes}

\author{
  \textbf{Xun Wang} \quad
  \textbf{Bihe Zhao} \quad
  \textbf{Michael Backes} \\
  \textbf{Franziska Boenisch} \quad
  \textbf{Adam Dziedzic} \\[0.5em]
  \normalfont CISPA Helmholtz Center for Information Security \\
  \normalfont\small
  \texttt{\{xun.wang,bihe.zhao,backes,boenisch,adam.dziedzic\}@cispa.de}
}

\begin{document}
\maketitle

\begin{abstract}
Commercial LLM APIs advertise a specific foundation model, but the served backbone may be silently substituted, quantized, or wrapped, for example to save deployment costs. All existing audits decide backbone identity from the text-output channel, which is structurally fragile for agentic APIs because modern serving stacks (OpenAI, Anthropic, Gemini, Cloudflare Workers AI, LangGraph) discard text and expose only structured actions when the model calls a tool, and \CR{provider-injected system prompts can distort text distributions enough that text-channel tests falsely accuse honest providers of substituting the claimed model}. We observe that recent agentic post-training internalizes tool-use directly into the weights, opening a new audit channel that the serving stack still exposes and that is \CR{largely} invariant to deployment context. We introduce \oursfull\ (\ours), the first action-based identity audit for agentic LLM APIs: \ours\ fingerprints a deployed model through its categorical tool-call distribution and decides identity via an MMD permutation test. \ours\ catches every substituted model ($100\%$ on $630$ evaluated checkpoint pairs), while holding the false-positive rate under system-prompt injection at $7\%$ (vs.\ $67\%$ for MET and $53\%$ for RUT). 
On third-party API endpoints, \ours's disagreements with MET are consistent with an independent token-count side-channel that detects \CR{provider-injected system prompts}.
\end{abstract}

\section{Introduction}
\label{sec:introduction}

\begin{figure}[t]
  \centering
  \includegraphics[width=\columnwidth]{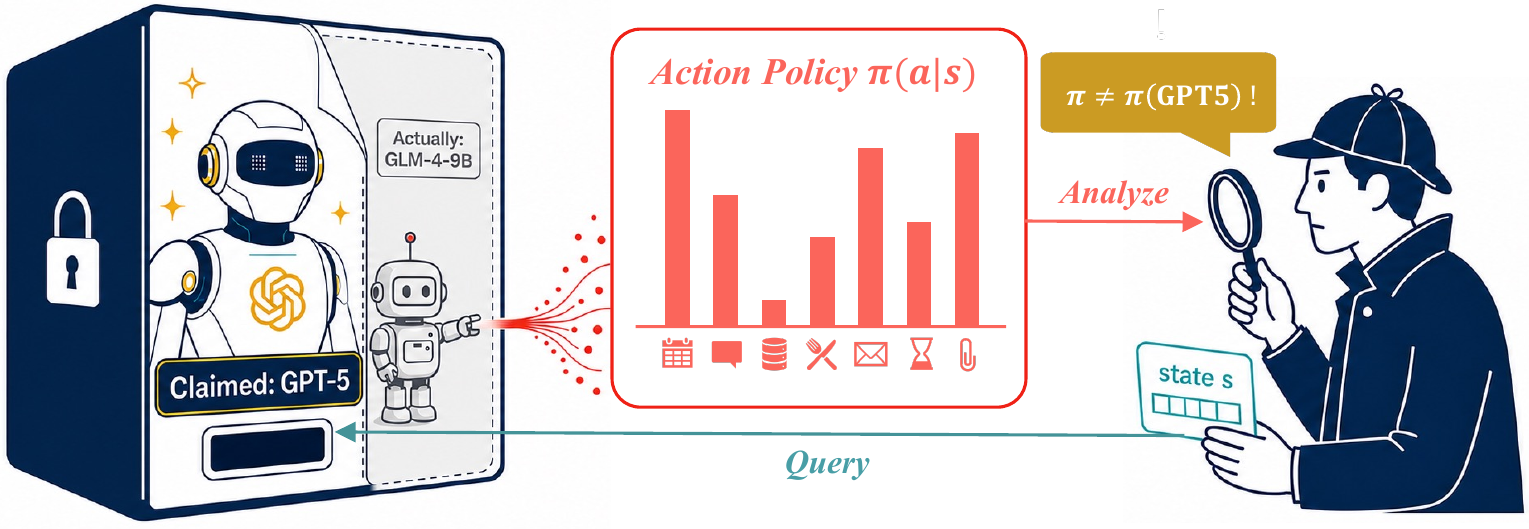}
  \caption{\CR{An auditor queries a suspect API endpoint with a probe state $s$; $a$ denotes the resulting tool-call action and $\pi(a \mid s)$ its conditional action policy. \ours\ compares this policy against a trusted reference and detects deviations from the claimed model, exposing a silent backbone substitution.}}
  \label{fig:teaser}
\end{figure}

Commercial LLMs are increasingly served through APIs that advertise a specific foundation model but operate as black boxes. The backbone behind an endpoint may be the advertised model, but can also be a quantized variant, a cheaper distilled substitute, or a wrapper routing to an open-weight alternative (Figure~\ref{fig:teaser}). Recent audits quantify the scale: \citet{Gao2025MET} flagged $11$ of $31$ commercial Llama endpoints as serving distributions different from their reference, and \citet{Zhang2026Shadow} surveyed $17$ third-party wrappers and reported $45.8\%$ endpoint-level identity failure, including wholesale backbone swaps such as GPT-5 to GLM-4-9B. Such discrepancies between the claimed and actually deployed models call for an auditing tool for probing what is behind the public API.

\CR{Most existing auditing methods, including MET~\citep{Gao2025MET} and RUT~\citep{Zhu2025RUT}, treat the model's natural-language output as its fingerprint, applying two-sample tests over text completions.} However, we observe two converging trends that are weakening natural-language completions as a reliable channel for agentic LLM API auditing. 
\CR{\textit{First,} production agentic serving stacks increasingly expose tool invocations as structured actions; in some configurations, they emit only the action, without the surrounding natural-language continuation~\citep{OpenAI2026FunctionCalling,Anthropic2026ToolUse}, which leaves text channel probes a limited space to operate on.}
\textit{Second}, providers routinely place their own system prompts and template chrome on top of the auditor's request~\citep{Zhang2026Shadow},
which shifts surface tokens without changing the underlying weights. As a consequence, a text-channel probe rejects identity and falsely accuses an honest provider of substituting the claimed model, only due to the different system prompts that the API providers adopt.

On top of the pure text output, tool-use behavior is now trained directly into model weights of agentic LLMs via dedicated post-training pipelines~\citep{Schick2023Toolformer, Patil2023Gorilla, Yang2025Qwen3, Team2025KimiK2}.
The resulting tool-call policy is a behavioral signature of the model that the serving stack still fully exposes, even when text is not fully exposed. 
In addition, the tool choice behavior is robust to the changes on the prompt surface: small prompt-level changes rarely flip \emph{which} tool the model invokes in a given situation, so action-level decisions are robust to exactly the deployment-context modifications that move text distributions.

Motivated by the above observation, we propose \oursfull\ (\ours), the first action-based identity audit for agentic LLM APIs. Given a claimed model and a suspect endpoint, \ours\ places both in a small number of tool-use situations and prompts the model to select from a given tool pool for solving the task.
The empirical distribution over these tool-call choices is the model's fingerprint, and we build a Maximum Mean Discrepancy (MMD) test on the collected fingerprint to decide whether the deployed model behind the API exhibits the expected policy. Our proposed \ours can operate even when text is stripped, and it is \CR{largely} invariant to the system prompts and templates providers inject around the model.

Our contributions are summarized as follows:
\begin{enumerate}
  \item We identify structural failures of text-channel API audits, observing that provider-side system prompts cause text-channel tests to falsely accuse honest providers of serving a substituted model.
  \item We propose \oursfull\ (\ours), the first action-based identity audit for agentic LLM APIs, which fingerprints a deployed model through its categorical tool-call distribution on probe states and decides identity via an MMD test.
  \item Our experimental results demonstrate that \ours\ accurately catches model substitution with $100\%$ rejection on $630$ evaluated checkpoint pairs. More importantly, \ours keeps the false-positive rate under system-prompt injection at $7\%$ versus $67\%$ for MET and $53\%$ for RUT.
\end{enumerate}

\section{Background}
\label{sec:background}

\CR{\textbf{Model Equality Testing }(MET)~\citep{Gao2025MET} formalizes the API audit problem as a two-sample test on completion distributions, deciding $H_0: P \equiv Q$ between a suspect endpoint and an authoritative reference using a Maximum Mean Discrepancy (MMD) statistic with a Hamming kernel. Applied to $31$ commercial Llama endpoints, MET flagged $11$ as silently drifted. RUT~\citep{Zhu2025RUT} instead constructs randomized ranks from log-rank response scores and tests them for uniformity with a Cram\'{e}r--von Mises goodness-of-fit test, reporting greater power than MET against the evaluated substitutions.}
\CR{Separately, \citet{Zhang2026Shadow} surveyed $17$ commercial wrapper APIs reselling frontier models (GPT-5, Gemini-2.5, DeepSeek-R1) and evaluated a $24$-endpoint subset ($3$ providers $\times$ $8$ models) with LLMmap, supplemented by MET analyses, finding $45.8\%$ identity-verification failure and documenting wholesale backbone swaps such as GPT-5 to GLM-4-9B.}
All three rely on the raw natural-language completion, which is an increasingly unreliable signal for agentic APIs since production serving stacks return only the structured action and provider-side prompts shift surface tokens.

\textbf{Black-box LLM Fingerprinting} also pursues black-box LLM identification under different threat models. \emph{Untargeted classification} methods identify an endpoint's backbone from a known set of candidates: LLMmap~\citep{Pasquini2025LLMmap} classifies $42$ LLM versions from $8$ queries, DuFFin~\citep{Yan2025DuFFin} couples a trigger with a knowledge fingerprint, and CoTSRF~\citep{Ren2025CoTSRF} exploits chain-of-thought contrast, which is inapplicable to the $82.8\%$ of commercial agentic platforms that expose tool-call traces while concealing internal reasoning~\citep{Wang2026AgentWM}.
Further variants rely on evolutionary query design~\citep{Iourovitski2024HideSeek} and stylometric artifacts~\citep{McGovern2025LeavingFingerprints}. \CR{\citet{Yang2026Unveiling} instead analyzes model-specific output subspaces without injecting a fingerprint.} \emph{Targeted verification} methods verify a specific identity via crafted trigger-response pairs: TRAP~\citep{Gubri2024TRAP}, ProFLingo~\citep{Jin2024ProFLingo}, RAFP~\citep{Tsai2025RoFL}, and SRAF~\citep{Wang2025RAPSM} optimize adversarial prompts, while injection-based schemes fine-tune signatures into the model via trigger-response backdoors~\citep{Xu2024Instructional, Yamabe2024MergePrint, Shao2025FITPrint, Shao2025Explanation}, scalable multi-fingerprint embedding~\citep{Nasery2025Scalable}
\emph{Active watermarking} on the agentic surface is pursued by AGENTWM~\citep{Wang2026AgentWM}, which injects user-specific biases into action sequences via five tool-equivalence schemes to enable ownership claims. 
However, AGENTWM is aimed to solve active ownership protection and requires model modification, while our method focuses on solving the passive equality verification on unmodified deployments.

\textbf{LLM Agentic Behavior} is trained directly into model weights through dedicated post-training pipelines~\citep{Schick2023Toolformer, Patil2023Gorilla, Team2025KimiK2, Yang2025Qwen3, Feng2025ReTool} rather than scaffolded by inference-time templates~\citep{Yao2022ReAct, Shinn2023Reflexion}. Different labs train on different data and recipes, and the resulting policies diverge in stable, reproducible ways: BFCL~\citep{Patil2025BFCL} and $\tau^2$-bench~\citep{Barres2025Tau2} report substantial per-model variation on identical function-call inputs, and BiasBusters~\citep{Blankenstein2026BiasBusters} shows each model carries a \emph{systematic, model-specific bias} over which tool it selects when several are functionally adequate. \ours\ is designed to capture this bias in the tool-use behavior through a single tool-selection probe instantiated across $K$ templates spanning general-tool and MCP server naming conventions (\S\ref{sec:probe}).

\section{Method}
\label{sec:method}

\textbf{\ours} models API provider accountability as a two-sample equality test on the \emph{tool-selection policy}: which tool the model picks when offered functionally-equivalent alternatives. Given black-box query access to a \emph{suspect} endpoint claiming to serve model~$M$ and \emph{reference} access to an authoritative instance of~$M$, the framework produces a binary accept/reject decision at a controlled false-positive rate. \CR{For open-weight models, the auditor can run the claimed checkpoint locally. For closed-source models, the vendor's official API can serve as the reference when auditing resellers, proxies, wrappers, or other third-party endpoints that claim to provide that model.} Three components make this operational: (i)~a tool-preference probe that queries the model on $K$ templates with functionally-equivalent tool options (\S\ref{sec:probe}), (ii)~a one-hot fingerprint with a squared $L_2$ test statistic (\S\ref{sec:statistic}), and (iii)~a permutation-calibrated null with a decision rule (\S\ref{sec:decision}).

\subsection{Problem Formulation}
\label{sec:formulation}

Each probe template $k \in [K]$ presents the model with functionally-equivalent tool options and a neutral user request. The model's tool selection is classified into an option set $\mathcal{O}_k$ (typically $|\mathcal{O}_k| \in [4{,}6]$). Let $\refpolicy$ and $\suspolicy$ denote the reference and suspect tool-selection distributions over these option sets. We formulate the following null hypothesis:
\begin{equation}
  H_0 : \refpolicy = \suspolicy,
  \label{eq:hypothesis}
\end{equation}
against the alternative that the distributions differ, at false-positive rate $\alpha = 0.05$. The test aggregates across all $K$ templates into a single statistic (\S\ref{sec:statistic}), calibrated via permutation (\S\ref{sec:decision}).

Because the test operates on categorical tool selections rather than raw text, it is expected to be more robust to deployment-context modifications: a provider-side system prompt shifts the literal output tokens but does not need to change which tool the model calls. \CR{We test this hypothesis in \S\ref{sec:exp-hidden-prompt}.}

\subsection{Tool-Preference Probe}
\label{sec:probe}

The probe exploits a single behavioral signal: when presented with multiple functionally-equivalent tools and a neutral request, different model lineages exhibit distinct tool-selection preferences inherited from their pre-training data and tool-use fine-tuning. The probe consists of $K$ templates (we use $K = 20$ in our primary evaluation). Each template specifies a set of tool schemas offering functionally-equivalent alternatives for the same operation (e.g., three calendar-creation tools with different names drawn from different provider conventions), a neutral user request, and a categorical option set $\mathcal{O}_k$ used by the auditor to classify the model's tool selection. The model never sees $\mathcal{O}_k$, and classification acts on the tool name in the model's unconstrained generation.

The $K$ templates span two complementary domains drawn as a single pool: \emph{general tools} (synthetic redundancy across $15$ everyday task categories) and \emph{MCP servers} (equivalent operations from different MCP server families~\citep{Anthropic2024MCP}). Both domains use tool surfaces that models plausibly encountered during training, so the probe reads the pre-trained tool-selection prior rather than an arbitrary classification task. Full template sourcing details are in Appendix~\ref{sec:probe-examples}.

Two mandatory controls apply to every template. First, tool order in the schema is randomized per sample with a deterministic seed, eliminating position bias. Second, tool descriptions within a template are length-matched and structurally parallel, preventing detail-driven selection. Prompts use neutral recipients and tool-agnostic task wording (``schedule a meeting'' rather than ``use Google Calendar'') to avoid domain priming.

\paragraph{Action classification.} \CR{The auditor maps each returned tool name to the corresponding option. Responses without a tool invocation and responses with malformed tool-call syntax are assigned separate outcomes. These outcomes are included in every option set. Because classification uses the tool invocation returned by the API, it is deterministic and remains applicable when no natural-language response is available.}

\subsection{Test Statistic}
\label{sec:statistic}

We build on the Maximum Mean Discrepancy (MMD) framework~\citep{Gretton2012}. Given a kernel $\kappa$ and two distributions $P,Q$, the MMD maps each distribution to its kernel mean embedding $\mu_P = \E_{x \sim P}[\kappa(x, \cdot)]$ in the reproducing kernel Hilbert space $\mathcal{H}_\kappa$ and measures their distance: $\MMD_\kappa(P,Q) = \|\mu_P - \mu_Q\|_{\mathcal{H}_\kappa}$. When $\kappa$ is \emph{characteristic}, $\MMD_\kappa(P,Q) = 0$ if and only if $P = Q$, making it a valid metric for two-sample testing.

Let $\mathcal{Z} = \{(k,a) : k \in [K],\, a \in \mathcal{O}_k\}$ denote the joint template-option alphabet. We use the delta kernel on $\mathcal{Z}$:
\begin{equation}
  \kappa\bigl((k,a),\,(k',a')\bigr) \;=\; \indicator\{k=k'\}\cdot\indicator\{a=a'\},
  \label{eq:kernel}
\end{equation}
which returns $1$ if and only if two samples originate from the same template and received the same classified option. The delta kernel on a finite alphabet is characteristic, so any difference in the tool-selection distribution is detectable with enough samples.

\paragraph{Fingerprint.} Each observed action is one-hot encoded over $\mathcal{Z}$, giving total dimension $D = \sum_{k=1}^{K} |\mathcal{O}_k|$. Collecting $N$ samples per template yields $N_{\text{total}} = K \cdot N$ rows per endpoint. The \emph{empirical fingerprint} is the column mean:
\begin{equation}
  \fingerprint \;=\; \frac{1}{N_{\text{total}}} \sum_{i=1}^{N_{\text{total}}} \mathbf{z}_i \;\in\; \R^D,
  \label{eq:fingerprint}
\end{equation}
i.e., the empirical distribution over tool selections within each template.

\begin{figure*}[t]
  \centering
  \resizebox{\textwidth}{!}{%
  \begin{tikzpicture}[
    lbl/.style={font=\footnotesize},
    sublbl/.style={font=\scriptsize, black!75},
    cell/.style={draw, minimum width=3.2mm, minimum height=3.2mm, inner sep=0pt,
                 font=\fontsize{4.5}{5.5}\selectfont},
    hot/.style={cell, fill=blue!28, font=\fontsize{4.5}{5.5}\selectfont\bfseries},
    zero/.style={cell, text=black!55, font=\fontsize{4.5}{5.5}\selectfont},
    blk/.style={draw, rounded corners=1pt, semithick},
    ar/.style={->, >=latex, semithick, black!65},
    optlbl/.style={font=\fontsize{4.5}{5.5}\selectfont, black!75, rotate=50, anchor=south west},
    steplbl/.style={font=\scriptsize\bfseries, black!85},
  ]
  \usetikzlibrary{positioning, decorations.pathreplacing}



  \begin{scope}[shift={(0,-0.50)}]
    \draw[rounded corners=3pt, fill=blue!4, draw=blue!35, line width=0.4pt]
      (0, -0.05) rectangle (4.8, -1.15);
    \fill[blue!55, rounded corners=1.5pt] (0.04, -0.10) rectangle (0.14, -1.10);
    \node[lbl, anchor=north west] at (0.25, -0.08)
      {$k_1$: ``Send an email''};
    \node[sublbl, anchor=north west] at (0.27, -0.42)
      {Tools offered: \textsf{gmail}, \textsf{outlook}, \textsf{yahoo}};
    \node[sublbl, anchor=north west] at (0.27, -0.78)
      {$\mathcal{O}_{k_1}\!=\!\{\textsf{gmail},\textsf{outlook},\textsf{yahoo},\textsf{\_no\_call}\}$};

    \draw[rounded corners=3pt, fill=blue!4, draw=blue!35, line width=0.4pt]
      (0, -1.65) rectangle (4.8, -2.75);
    \fill[blue!55, rounded corners=1.5pt] (0.04, -1.70) rectangle (0.14, -2.70);
    \node[lbl, anchor=north west] at (0.25, -1.68)
      {$k_2$: ``Schedule a meeting''};
    \node[sublbl, anchor=north west] at (0.27, -2.02)
      {Tools offered: \textsf{gcal}, \textsf{zoom}, \textsf{calendly}};
    \node[sublbl, anchor=north west] at (0.27, -2.38)
      {$\mathcal{O}_{k_2}\!=\!\{\textsf{gcal},\textsf{zoom},\textsf{calendly},\textsf{\_no\_call}\}$};

    \node[sublbl, anchor=north west] at (0.08, -3.05)
      {Query $N\times$; record the model's tool pick.};
  \end{scope}

  \begin{scope}[shift={(6.4, 0.05)}]
    \def\cw{0.36}
    \foreach \i/\t in {0/gmail, 1/outlk, 2/yahoo, 3/no\_cll, 4.3/gcal, 5.3/zoom, 6.3/cldly, 7.3/no\_cll}{
      \node[optlbl] at (\i*\cw-0.06, -0.25) {\textsf{\t}};}

    \def\ry{-0.72}
    \foreach \r/\lab in {0/{$k_1$ \#1}, 1/{$k_1$ \#2}, 2/{$k_1$ \#3}, 3/{$k_2$ \#1}, 4/{$k_2$ \#2}, 5/{$k_2$ \#3}}{
      \node[sublbl, anchor=east] at (-0.12, \ry-\r*0.36) {\lab};
    }
    \foreach \i/\v/\s in {0/0/zero,1/1/hot,2/0/zero,3/0/zero, 4.3/0/zero,5.3/0/zero,6.3/0/zero,7.3/0/zero}{
      \node[\s] at (\i*\cw, \ry) {\v};}
    \foreach \i/\v/\s in {0/1/hot,1/0/zero,2/0/zero,3/0/zero, 4.3/0/zero,5.3/0/zero,6.3/0/zero,7.3/0/zero}{
      \node[\s] at (\i*\cw, \ry-0.36) {\v};}
    \foreach \i/\v/\s in {0/0/zero,1/1/hot,2/0/zero,3/0/zero, 4.3/0/zero,5.3/0/zero,6.3/0/zero,7.3/0/zero}{
      \node[\s] at (\i*\cw, \ry-0.72) {\v};}
    \foreach \i/\v/\s in {0/0/zero,1/0/zero,2/0/zero,3/0/zero, 4.3/1/hot,5.3/0/zero,6.3/0/zero,7.3/0/zero}{
      \node[\s] at (\i*\cw, \ry-1.08) {\v};}
    \foreach \i/\v/\s in {0/0/zero,1/0/zero,2/0/zero,3/0/zero, 4.3/0/zero,5.3/1/hot,6.3/0/zero,7.3/0/zero}{
      \node[\s] at (\i*\cw, \ry-1.44) {\v};}
    \foreach \i/\v/\s in {0/0/zero,1/0/zero,2/0/zero,3/0/zero, 4.3/1/hot,5.3/0/zero,6.3/0/zero,7.3/0/zero}{
      \node[\s] at (\i*\cw, \ry-1.80) {\v};}

    \draw[blk, blue!40] (-0.18, \ry+0.18) rectangle (3*\cw+0.18, \ry-1.98);
    \draw[blk, blue!40] (4.3*\cw-0.18, \ry+0.18) rectangle (7.3*\cw+0.18, \ry-1.98);

    \def\arrowtop{-2.85}
    \def\arrowbot{-3.15}
    \draw[ar] (1.5*\cw, \arrowtop) -- (1.5*\cw, \arrowbot);
    \draw[ar] (5.8*\cw, \arrowtop) -- (5.8*\cw, \arrowbot);
    \node[sublbl, blue!70!black, anchor=west] at (1.5*\cw+0.07, {(\arrowtop+\arrowbot)/2}) {$k_1$};
    \node[sublbl, blue!70!black, anchor=west] at (5.8*\cw+0.07, {(\arrowtop+\arrowbot)/2}) {$k_2$};
    \node[sublbl, anchor=west] at (7.3*\cw+0.05, {(\arrowtop+\arrowbot)/2}) {col.\ mean};

    \def\fy{-3.45}
    \node[lbl, anchor=east, blue!70!black] at (-0.15, \fy) {$\fingerprint =$};
    \foreach \i/\v in {0/.17, 1/.33, 2/.00, 3/.00, 4.3/.33, 5.3/.17, 6.3/.00, 7.3/.00}{
      \node[cell, fill=blue!12, font=\fontsize{4}{5}\selectfont] at (\i*\cw, \fy) {\v};}
    \draw[blk, blue!25] (-0.18, \fy-0.18) rectangle (3*\cw+0.18, \fy+0.18);
    \draw[blk, blue!25] (4.3*\cw-0.18, \fy-0.18) rectangle (7.3*\cw+0.18, \fy+0.18);
    \node[sublbl, anchor=west] at (7.3*\cw+0.28, \fy) {$\!\in \R^D$};
  \end{scope}

  \begin{scope}[shift={(11.0, 0.30)}, scale=1.2, every node/.append style={transform shape}]
    \node[sublbl] at (0.675, -0.55) {$k_1$};
    \node[sublbl] at (2.225, -0.55) {$k_2$};

    \node[sublbl, blue!70!black] at (1.45, -0.93)
      {Reference $\fingerprint^R$};
    \fill[blue!25] (0, -1.55) rectangle (0.3, -1.10);
    \fill[blue!25] (0.35, -1.55) rectangle (0.65, -1.25);
    \fill[blue!25] (0.70, -1.55) rectangle (1.00, -1.50);
    \fill[blue!25] (1.05, -1.55) rectangle (1.35, -1.50);
    \fill[blue!25] (1.55, -1.55) rectangle (1.85, -1.20);
    \fill[blue!25] (1.90, -1.55) rectangle (2.20, -1.35);
    \fill[blue!25] (2.25, -1.55) rectangle (2.55, -1.45);
    \fill[blue!25] (2.60, -1.55) rectangle (2.90, -1.50);
    \draw[gray!50] (0, -1.55) -- (2.90, -1.55);

    \node[sublbl, red!70!black] at (1.45, -1.88)
      {Suspect $\fingerprint^S$};
    \fill[red!25] (0, -2.50) rectangle (0.3, -2.30);
    \fill[red!25] (0.35, -2.50) rectangle (0.65, -2.45);
    \fill[red!25] (0.70, -2.50) rectangle (1.00, -2.10);
    \fill[red!25] (1.05, -2.50) rectangle (1.35, -2.45);
    \fill[red!25] (1.55, -2.50) rectangle (1.85, -2.30);
    \fill[red!25] (1.90, -2.50) rectangle (2.20, -2.05);
    \fill[red!25] (2.25, -2.50) rectangle (2.55, -2.25);
    \fill[red!25] (2.60, -2.50) rectangle (2.90, -2.45);
    \draw[gray!50] (0, -2.50) -- (2.90, -2.50);

    \draw[gray!40, dashed] (1.45, -1.05) -- (1.45, -2.55);

    \node[lbl, anchor=north] at (1.45, -2.85)
      {$T = {\textstyle\sum_k} \|\fingerprint^R_k {-} \fingerprint^S_k\|^2_2$};
  \end{scope}

  \begin{scope}[shift={(15.6, 0.26)}]
    \def\bw{0.42}
    \def\base{-3.30}
    \def\hscale{1.55}
    \fill[gray!22] (0*\bw, \base) rectangle (1*\bw, \base+0.20*\hscale);
    \fill[gray!22] (1*\bw, \base) rectangle (2*\bw, \base+0.50*\hscale);
    \fill[gray!22] (2*\bw, \base) rectangle (3*\bw, \base+0.85*\hscale);
    \fill[gray!22] (3*\bw, \base) rectangle (4*\bw, \base+1.20*\hscale);
    \fill[gray!22] (4*\bw, \base) rectangle (5*\bw, \base+1.35*\hscale);
    \fill[gray!22] (5*\bw, \base) rectangle (6*\bw, \base+1.05*\hscale);
    \fill[gray!22] (6*\bw, \base) rectangle (7*\bw, \base+0.65*\hscale);
    \fill[gray!22] (7*\bw, \base) rectangle (8*\bw, \base+0.30*\hscale);
    \fill[gray!22] (8*\bw, \base) rectangle (9*\bw, \base+0.10*\hscale);
    \draw[gray!55] (-0.05, \base) -- (9*\bw+0.05, \base);

    \draw[dashed, orange!80!black, semithick] (7*\bw, \base-0.08) -- (7*\bw, \base+1.60*\hscale);
    \node[sublbl, orange!80!black, anchor=south, align=center] at (7*\bw, \base+1.60*\hscale)
      {reject\\[-1pt]threshold};

    \draw[red!70!black, thick] (8.5*\bw, \base-0.08) -- (8.5*\bw, \base+1.50*\hscale);
    \node[font=\scriptsize, red!70!black, anchor=south] at (8.5*\bw, \base+1.50*\hscale) {$T_{\text{obs}}$};

    \draw[<-, >=latex, red!70!black, semithick] (8.75*\bw, \base+0.80*\hscale) -- (8.75*\bw+0.7, \base+0.80*\hscale);
    \node[font=\scriptsize\bfseries, red!70!black, anchor=west] at (8.75*\bw+0.7, \base+0.80*\hscale) {Reject $H_0$};

    \node[sublbl, anchor=north] at (4.5*\bw, \base-0.08) {$T$};
    \node[sublbl, rotate=90] at (-0.30, \base+0.75*\hscale) {count};
    \node[sublbl, anchor=south] at (4.5*\bw, \base+1.85*\hscale)
      {Null distribution of $T$};
    \node[sublbl, anchor=north, align=center] at (4.5*\bw, \base-0.28)
      {pool responses from both endpoints,\\[-1pt]randomly re-split, recompute $T$};
  \end{scope}

  \node[steplbl, anchor=base] at (2.40,  -4.35) {(a) Probe templates};
  \node[steplbl, anchor=base] at (7.71,  -4.35) {(b) One-hot encoding};
  \node[steplbl, anchor=base] at (12.74, -4.35) {(c) Test statistic};
  \node[steplbl, anchor=base] at (17.49, -4.35) {(d) Permutation null};

  \end{tikzpicture}%
  }
  \caption{\textbf{Our \ours pipeline.} (a)~Each template offers the model functionally-equivalent tools (e.g., gmail/outlook/yahoo) and we query $N$ times. (b)~Each query becomes a one-hot row marking which tool was chosen and the column average is the fingerprint vector $\fingerprint$. (c)~The test statistic $T$ is the squared $L_2$ distance between reference and suspect fingerprints. (d)~A permutation null calibrates the test. We reject if the observed $T$ exceeds the $\alpha{=}0.05$ threshold.}
  \label{fig:pipeline}
\end{figure*}
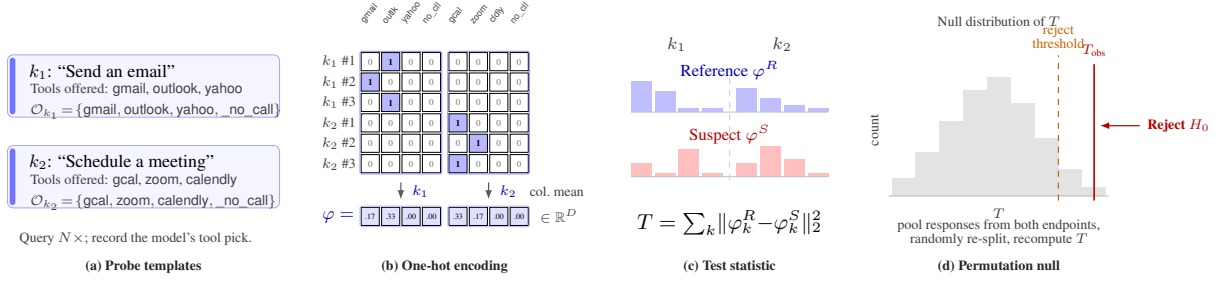

\paragraph{Test statistic.} The squared population MMD, $\MMD^2_\kappa(\refpolicy, \suspolicy)$, measures the distributional gap between the two policies in the kernel's feature space. Under the delta kernel, its empirical estimator $\hMMD^2_\kappa$ reduces to the squared $L_2$ distance between fingerprints:
\begin{equation}
  T \;=\; \bigl\|\fingerprint^{R} - \fingerprint^{S}\bigr\|^2_2 \;=\; \hMMD^2_\kappa\!\bigl(\mathbf{Z}^R, \mathbf{Z}^S\bigr).
  \label{eq:test-stat}
\end{equation}
Because the one-hot encoding is block-diagonal (each row has nonzero entries in exactly one template's block), the statistic decomposes by template:
\begin{equation}
  T \;=\; \sum_{k=1}^{K} \bigl\|\fingerprint^{R}_{k} - \fingerprint^{S}_{k}\bigr\|^2_2,
  \label{eq:block-decomp}
\end{equation}
where $\fingerprint^{R}_{k}$ denotes the restriction of $\fingerprint^{R}$ to template~$k$'s coordinates. This per-template breakdown serves as an interpretability diagnostic (\S\ref{sec:decision}).

\subsection{Null Calibration and Decision}
\label{sec:decision}

\paragraph{Permutation null.} 
Under $H_0$, endpoint labels are exchangeable within each template,
assuming queries are independent conditional on the template. For each
template, we pool its $2N$ rows, shuffle endpoint labels uniformly
within that template, and resplit them into groups of size $N$. We
combine the resulting groups across templates, recompute $T$, and repeat
this procedure $B=1{,}000$ times to obtain the empirical null
distribution $\{T^{(r)}\}_{r=1}^{B}$.

\paragraph{Decision rule.} We reject $H_0$ at level $\alpha = 0.05$ if $T$ exceeds the $(1-\alpha)$-quantile of the null distribution (Algorithm~\ref{alg:agentprov}). Type-I error is controlled by the exchangeability construction, regardless of $N$, $K$, or the option-set sizes. The per-template terms $\|\fingerprint^{R}_{k} - \fingerprint^{S}_{k}\|^2_2$ from Eq.~\ref{eq:block-decomp} identify which tool-selection scenarios contributed most to any rejection. This diagnostic is descriptive and does not affect the aggregate's Type-I control.

\begin{algorithm}[t]
  \caption{\ours{} tool-selection policy equality test}
  \label{alg:agentprov}
  \begin{algorithmic}[1]
    \Require Reference $\refpolicy$, suspect $\suspolicy$, probe templates $\{\mathcal{T}_k\}_{k=1}^{K}$, samples per template $N$, permutations $B$, level $\alpha$.
    \For{each template $k \in [K]$}
      \State Sample $N$ tool calls from $\refpolicy$ and $\suspolicy$ on template $\mathcal{T}_k$; classify each into option $a \in \mathcal{O}_k$.
      \State Encode each classified action as a one-hot row $\mathbf{z}_i \in \{0,1\}^D$.
    \EndFor
    \State Stack all rows into matrices $\mathbf{Z}^R, \mathbf{Z}^S \in \{0,1\}^{N_{\text{total}} \times D}$.
    \State Compute fingerprints $\fingerprint^R = \mathrm{colmean}(\mathbf{Z}^R)$, $\fingerprint^S = \mathrm{colmean}(\mathbf{Z}^S)$.
    \State Compute observed $T = \|\fingerprint^R - \fingerprint^S\|^2_2$.
    \For{$r = 1, \dots, B$}
      \State Shuffle uniformly within each template and resplit preserving sample sizes.
      \State Recompute $T^{(r)}$ on the resplit pair.
    \EndFor
    \State $q_{1-\alpha} \leftarrow$ empirical $(1-\alpha)$-quantile of $\{T^{(r)}\}_{r=1}^{B}$.
    \State \Return \textsc{Reject} if $T > q_{1-\alpha}$, else \textsc{Accept}.
  \end{algorithmic}
\end{algorithm}

\section{Evaluation}
\label{sec:evaluation}

\CR{We evaluate \ours across scenarios relevant to agentic-API accountability audits:} distinguishing different models (\S\ref{sec:exp-controlled}), separating identity drift from deployment-context drift (\S\ref{sec:exp-hidden-prompt}), verifying third-party API providers (\S\ref{sec:exp-api}), and characterizing efficiency (\S\ref{sec:exp-ablations})\CR{, tool-name and schema robustness (\S\ref{sec:exp-probe-surface}), and robustness to adaptive providers (\S\ref{sec:exp-adaptive})}.

\subsection{Setup}
\label{sec:exp-setup}

\paragraph{Models.} Our pool spans open-weight configurations across $13$ families ($0.5$B--$14$B parameters): $36$ base checkpoints for local identity tests, $5$ local checkpoints tested under injected system prompts, and \CR{$9$} third-party API deployments compared against their local reference. \CR{We additionally evaluate $9$ closed-source OpenAI models through the official API and a third-party reseller. All model names are listed in Appendix Table~\ref{tab:models}.}

\paragraph{Baselines.} We compare \ours\ against MET~\citep{Gao2025MET} (MMD with Hamming kernel on token-string completions), RUT~\citep{Zhu2025RUT} (Cram\'{e}r--von Mises on reference-model log-rank scores), \CR{and our two-sample adaptation of LLMmap~\citep{Pasquini2025LLMmap} (MMD$^2$ on E5 response embeddings)}. RUT is grey-box: it requires logprobs from the reference, so it applies only when the reference is run locally or its API returns logits.

\paragraph{Protocol.} \CR{MET and RUT follow their published collection and decision protocols.} Method-specific parameters (sample sizes, prompts, kernel and null-calibration settings) are in Appendix~\ref{sec:protocol-details}. All tests calibrate at significance $\alpha = 0.05$ via the method's own null. The joint $K \times N$ budget sweep is in \S\ref{sec:exp-ablations}.

\subsection{\CR{Model Identity Test}}
\label{sec:exp-controlled}

We first verify that \ours\ separates distinct checkpoints as reliably as the text-channel baselines. \CR{At $K = 20$ templates of the tool-selection probe and $N = 50$ samples per template, all three methods reject all $630$ distinct-checkpoint pairs at $\alpha = 0.05$ ($100\%$, Table~\ref{tab:method-comparison}).} On the $36$ same-model self-pairs, both \ours\ and MET never falsely reject, while RUT falsely rejects $2$ models.

\begin{table}[t]
\centering
\small
\setlength{\tabcolsep}{4pt}
\begin{tabular}{lcccc}
\toprule
\textbf{Category} & $n$ & MET & RUT & \ours \\
\midrule
Cross-pairs        & $630$ & $100\%$ & $100\%$ & $100\%$ \\
Self-pairs &  $36$ & $0/36$  & $2/36$ & $0/36$ \\
\bottomrule
\end{tabular}
\caption{Identity tests on $36$ checkpoints at $\alpha=0.05$. \ours\ uses the tool-selection probe at $K=20$ templates, $N=50$ samples per template. All methods reject every cross-pair; on self-pairs, MET and \ours\ never falsely reject while RUT does so on $2$ models.}
\label{tab:method-comparison}
\end{table}

\begin{figure}[t]
  \centering
  \includegraphics[width=\columnwidth]{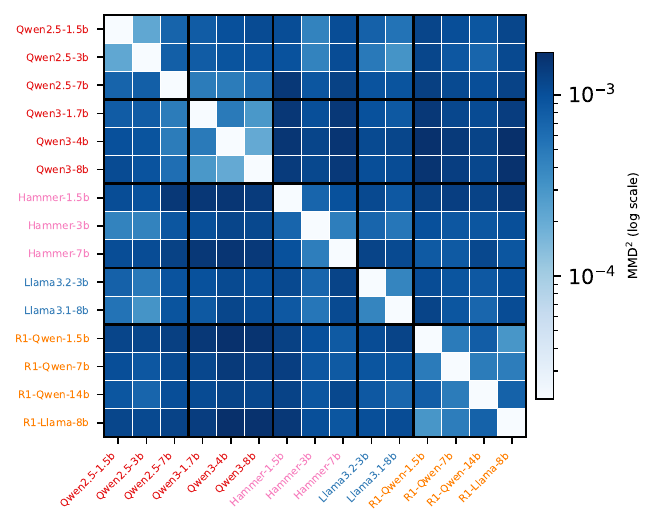}
  \caption{Pairwise MMD$^2$ among $15$ representative models (log scale), \ours's tool-selection probe at $K{=}20$, $N{=}50$. Every off-diagonal pair rejects at $\alpha{=}0.05$. Within-family distances are visibly smaller for Qwen$2.5$, Qwen$3$, Hammer-$2.1$, and Llama; R1-Distill exhibits large within-family variance. The Qwen$2.5$--Hammer cross-block is lighter than other cross-family regions, reflecting their shared pre-training base. Full $36{\times}36$ matrix in Appendix~\ref{app:full-heatmap}.}
  \label{fig:pairwise}
\end{figure}

Beyond the binary decision,
the test statistic $T$ provides a continuous distance over the action distribution. Figure~\ref{fig:pairwise} shows the pairwise matrix on a $15$-model representative subset: same-family models cluster (Qwen$2.5$ across sizes, Llama-$3$ across sizes) and shared training lineage carries across architectures (R1-Distill variants cluster regardless of base). The full $36 \times 36$ matrix is in Appendix~\ref{app:full-heatmap}.

\paragraph{\CR{Frontier models.}} \CR{We extend the controlled identity tests to the nine closed-source OpenAI models listed in Appendix Table~\ref{tab:models}, accessed through the official API. Across independent same-model samples, \ours accepts all $9/9$ self-comparisons, rejects all $36/36$ distinct-model pairs, and rejects all $324/324$ comparisons between the nine OpenAI models and the $36$ local checkpoints. These results extend the local-checkpoint findings to closed-source models when the official endpoint provides the authoritative reference.}

\paragraph{Family cohesion and post-training drift.} Table~\ref{tab:intra-cross} quantifies the heatmap's visual structure: per family, the mean within-family MMD$^2$ versus the mean between-family MMD$^2$. \CR{Seven of the eight multi-member families} cluster tightly (cross/intra ratios $1.8$--$3.0\times$), confirming that the tool-selection prior is largely shared across sizes within a family. \CR{\emph{Hermes} is the sole multi-member exception} (ratio $0.8\times$) and applies the Hermes-3 SFT recipe to three different bases (Llama-$3.1$-$8$B, Llama-$3.2$-$3$B, Mistral-$7$B). Each Hermes endpoint stays closer to its own base ($T \in [7.1, 33.4]\times 10^{-4}$) than to its Hermes siblings, indicating that chat-persona SFT does not override the base's tool-selection bias. The fingerprint reads what was trained into the weights for tool use, not the SFT brand. \CR{The singleton \emph{Qwen3-Think} has no within-family ratio. Its reasoning-mode post-training on a shared base nonetheless produces a large shift:} \texttt{qwen3-4b-thinking-2507} sits $\sim$$14\times$ farther from the other four Qwen$3$ checkpoints than they sit from each other, showing that reasoning-mode SFT substantially rewires the action policy and that \ours\ catches the drift.

\begin{table}[t]
  \centering
  \footnotesize
  \setlength{\tabcolsep}{4pt}
  \begin{tabular}{lrrr}
    \toprule
    Family & $T_{\text{intra}}$ & $T_{\text{cross}}$ & $T_{\text{c}} / T_{\text{i}}$ \\
    \midrule
    R1-Distill   & $5.43$ & $15.32$ & $2.8\times$ \\
    DeepSeek     & $5.57$ & $16.62$ & $3.0\times$ \\
    Qwen3        & $6.49$ & $17.28$ & $2.7\times$ \\
    Hammer-2.1   & $6.80$ & $15.99$ & $2.4\times$ \\
    Qwen2.5      & $7.29$ & $15.13$ & $2.1\times$ \\
    Llama        & $7.40$ & $15.43$ & $2.1\times$ \\
    Mistral      & $8.04$ & $14.54$ & $1.8\times$ \\
    Hermes       & $27.20$ & $22.89$ & $0.8\times$ \\
    Qwen3-Think  & -- & $107.04$ & -- \\
    \bottomrule
  \end{tabular}
  \caption{Family cohesion in tool-selection action space on the $36$-model set (\S\ref{sec:exp-controlled}), \ours's tool-selection probe at $K{=}20$, $N{=}50$. $T_{\text{intra}}$: mean MMD$^2$ across within-family pairs; $T_{\text{cross}}$: mean across between-family pairs (both in units of $10^{-4}$). Higher ratio $=$ tighter internal clustering. \emph{Qwen3-Think} has a single member (\texttt{qwen3-4b-thinking-2507}) so $T_{\text{intra}}$ is undefined.}
  \label{tab:intra-cross}
\end{table}

\subsection{Vulnerability to Hidden System Prompts}
\label{sec:exp-hidden-prompt}

Text-channel tests measure surface tokens: MET's Hamming kernel compares completions character-by-character, and RUT's log-rank score is a function of literal token positions under the reference distribution. Any change to the input that shifts the output token stream, even by one character, moves both statistics. In practice, providers routinely modify exactly this surface: in a token-count scan of $75$ OpenRouter endpoints, $44\%$ deviate from the model's official chat template, including endpoints that inject hidden system prompts (Appendix~\ref{sec:or-scan-method}). This combination, a test that responds to any token-level change paired with a deployment surface the auditor cannot observe, makes text-channel identity tests vulnerable. We confirm the vulnerability under controlled conditions.

\paragraph{Experimental design.} For each of $5$ open-weight checkpoints (listed in Appendix Table~\ref{tab:models}) we keep weights, decoding, and chat template fixed and vary only the system message across three conditions: an \emph{empty} message (a single whitespace, $1$ token), a \emph{short} boilerplate (``You are a helpful assistant.'', $\sim$$5$ tokens), and a \emph{long} provider-style customer-service prompt ($\sim$$50$ tokens). Each condition is tested against the same-model no-system-role baseline. As the weights are identical, every rejection is by construction a false positive on identity.

\paragraph{Result.} Table~\ref{tab:hidden-prompt-fpr} reports per-condition decisions. MET falsely rejects $3$/$5$ models even on the empty condition and $4$/$5$ on the long one. RUT climbs from $1$/$5$ to $5$/$5$ across the same range. \CR{Adapted LLMmap rejects $0/5$, $1/5$, and $2/5$ models under the empty, short, and long conditions, respectively.} \ours\ rejects $0$/$5$ on empty and short and $1$/$5$ on long, with the single rejection driven by Gemma-$3$-$1$B, whose tool-call format shifts under the longer persona-style prompt. Aggregated across all three conditions, \ours\ produces $1/15$ false positives ($7\%$) versus $10/15$ ($67\%$) for MET, $8/15$ ($53\%$) for RUT, \CR{and $3/15$ ($20\%$) for adapted LLMmap}. The asymmetry confirms the prediction of \S\ref{sec:statistic}: the delta kernel on categorical actions is insensitive to surface-text shifts that the Hamming and log-rank kernels respond to.

\begin{table*}[t]
\centering
\scriptsize
\setlength{\tabcolsep}{3pt}
\begin{tabular}{l ccc ccc ccc ccc}
\toprule
& \multicolumn{3}{c}{MET} & \multicolumn{3}{c}{RUT} & \multicolumn{3}{c}{\CR{Adapted LLMmap}} & \multicolumn{3}{c}{\ours} \\
\cmidrule(lr){2-4} \cmidrule(lr){5-7} \cmidrule(lr){8-10} \cmidrule(lr){11-13}
\textbf{Model} & empty & short & long & empty & short & long & \CR{empty} & \CR{short} & \CR{long} & empty & short & long \\
\midrule
\texttt{Llama-3.2-1B-Instruct}   & $\checkmark$ & $\checkmark$ & \textbf{$\times$} & $\checkmark$ & $\checkmark$ & \textbf{$\times$} & \CR{$\checkmark$} & \CR{$\checkmark$} & \CR{\textbf{$\times$}} & $\checkmark$ & $\checkmark$ & $\checkmark$ \\
\texttt{Qwen3-1.7B}              & \textbf{$\times$} & \textbf{$\times$} & \textbf{$\times$} & $\checkmark$ & \textbf{$\times$} & \textbf{$\times$} & \CR{$\checkmark$} & \CR{$\checkmark$} & \CR{$\checkmark$} & $\checkmark$ & $\checkmark$ & $\checkmark$ \\
\texttt{Qwen2.5-1.5B-Instruct}   & \textbf{$\times$} & \textbf{$\times$} & \textbf{$\times$} & $\checkmark$ & $\checkmark$ & \textbf{$\times$} & \CR{$\checkmark$} & \CR{$\checkmark$} & \CR{$\checkmark$} & $\checkmark$ & $\checkmark$ & $\checkmark$ \\
\texttt{Gemma-3-1B-it}           & \textbf{$\times$} & \textbf{$\times$} & \textbf{$\times$} & \textbf{$\times$} & \textbf{$\times$} & \textbf{$\times$} & \CR{$\checkmark$} & \CR{\textbf{$\times$}} & \CR{\textbf{$\times$}} & $\checkmark$ & $\checkmark$ & \textbf{$\times$} \\
\texttt{Hammer2.1-1.5B}          & $\checkmark$ & $\checkmark$ & $\checkmark$ & $\checkmark$ & $\checkmark$ & \textbf{$\times$} & \CR{$\checkmark$} & \CR{$\checkmark$} & \CR{$\checkmark$} & $\checkmark$ & $\checkmark$ & $\checkmark$ \\
\midrule
\textbf{FPR}                     & $3/5$ & $3/5$ & $4/5$ & $1/5$ & $2/5$ & $5/5$ & \CR{$0/5$} & \CR{$1/5$} & \CR{$2/5$} & \textbf{0/5} & \textbf{0/5} & \textbf{1/5} \\
\bottomrule
\end{tabular}
\caption{Per-model false-positive rate on the hidden-prompt control experiment (\S\ref{sec:exp-hidden-prompt}). Each cell reports the per-model decision at $\alpha=0.05$ when the system prompt is replaced by the column variant; weights are unchanged, so every $\times$ is a false positive on identity. The \emph{empty} condition is a single whitespace ($1$ token); \emph{short} is ``You are a helpful assistant.''~($\sim$$5$ tokens); \emph{long} is a $\sim$$50$-token provider-style customer-service prompt. \ours\ uses the tool-selection probe at $K=20$ templates, $N=50$ samples per template.}
\label{tab:hidden-prompt-fpr}
\end{table*}

The text-channel vulnerability appears at the shortest prompt we test: a single whitespace already flips MET on $3$/$5$ models. The effect is therefore not a corner case requiring crafted adversarial prompts. Any provider customization that adds a system role to the request can flip the decision on otherwise-identical samples. The long condition sits within the range observed in real provider deployments, and \S\ref{sec:exp-api} shows the same false-positive pattern on real third-party endpoints.

\subsection{API Provider Verification}
\label{sec:exp-api}

We apply \ours, MET, and RUT to a practical deployment scenario: verifying whether a third-party API endpoint serves the model it claims. The audit compares samples from a \emph{trusted reference} (local inference for open-weight models) against samples from a \emph{suspect endpoint} on OpenRouter. Since the true serving configuration is unknown to the auditor, no decision can be validated against ground truth. We therefore construct a side-channel signal to contextualize each method's output: a token count signal that compares the provider-reported \texttt{prompt\_tokens} against the auditor's locally computed count for the same messages (details in Appendix~\ref{sec:or-scan-method}). A persistent positive offset reveals content the provider injected but the auditor never specified, i.e., a hidden system prompt. This signal does not flag whether the model was substituted, but it does tell us whether the deployment context was modified, which \S\ref{sec:exp-hidden-prompt} showed text-channel tests are vulnerable to. \CR{Of the $9$ endpoints, $5$ show extra tokens indicative of hidden prompts and $4$ do not.} Table~\ref{tab:api} reports per-endpoint decisions. The key question is whether rejections on hidden-prompt endpoints reflect genuine model substitution or the deployment-context sensitivity.

\paragraph{Hidden-prompt endpoints.} MET rejects all $5$ endpoints flagged by the token-count scan. RUT rejects $4$ of $5$, accepting only the $+7$-token Gemma-$3$n-E$4$B. \ours\ accepts the four small-to-moderate injection cases (Llama-$3.2$-$1$B, Qwen$3$-$14$B, Llama-$3$-$8$B, each $\sim$$1$ extra token, and Gemma-$3$n-E$4$B at $+7$), consistent with the controlled finding of \S\ref{sec:exp-hidden-prompt} that injected system prompts at this size shift surface tokens but not tool selection. On Llama-$3.2$-$3$B ($+25$ extra tokens) all three methods reject, suggesting an injection large enough to also move the action distribution.

\paragraph{Endpoints without detected hidden prompts.} Among the \CR{four endpoints evaluated against matching local references} whose token counts match the official template, the methods disagree in informative ways. On Llama-$3.1$-$8$B, MET and \ours\ accept while RUT rejects, possibly due to precision differences between the served endpoint and the local reference. On Mistral-Nemo, all three methods reject, indicating a difference visible in both surface tokens and tool selection. The remaining two endpoints split: on Gemma-$3$-$4$B, RUT and \ours\ accept while MET rejects, suggesting a surface-text shift visible to MET's Hamming kernel but not to the action distribution or the log-rank score. On Qwen$3$-$8$B, MET and RUT both reject while \ours\ accepts, suggesting a text-channel deviation that does not extend to tool selection.

\begin{table}[t]
  \centering
  \small
  \setlength{\tabcolsep}{4pt}
  \begin{tabular}{l c c c}
    \toprule
    Endpoint                          & MET          & RUT          & \ours \\
    \midrule
    \multicolumn{4}{l}{\textit{With detected hidden prompt}} \\
    Llama-$3.2$-$3$B                   & rej & rej & rej \\
    Gemma-$3$n-E$4$B                   & rej & acc          & acc \\
    Llama-$3.2$-$1$B                   & rej & rej & acc \\
    Qwen$3$-$14$B                      & rej & rej & acc \\
    Llama-$3$-$8$B                     & rej & rej & acc \\
    \midrule
    \multicolumn{4}{l}{\textit{No detected hidden prompt}} \\
    Llama-$3.1$-$8$B                   & acc          & rej & acc \\
    Qwen$3$-$8$B                       & rej & rej & acc \\
    Gemma-$3$-$4$B                     & rej & acc          & acc \\
    Mistral-Nemo-Instruct-$2407$       & rej & rej & rej \\
    \bottomrule
  \end{tabular}
  \caption{Third-party API verification on \CR{$9$} OpenRouter endpoints (\S\ref{sec:exp-api}), grouped by whether the OpenRouter token-count scan (Appendix~\ref{sec:or-scan-method}) detected a provider-injected hidden system prompt. \CR{Across these endpoints, \ours\ accepts $7$, MET accepts $1$, and RUT accepts $2$.} Cells report decisions at $\alpha = 0.05$. \ours\ uses the tool-selection probe with $N=50$ samples per template; MET and RUT consume the same matched-provider API responses.}
  \label{tab:api}
\end{table}

\paragraph{Summary.} \CR{Across the nine endpoints, \ours\ accepts $7$, MET accepts $1$, and RUT accepts $2$.} The pattern is consistent with \ours\ being insensitive to deployment-context modifications that shift surface tokens (hidden prompts, template variants) while remaining sensitive to changes in the action distribution (model substitution, large prompt injections). Absent ground truth on serving configurations, these interpretations remain correlational.

\paragraph{\CR{OpenAI reseller control.}} \CR{We additionally compare nine OpenAI models served through OpenRouter with their matching official OpenAI endpoints. \ours accepts all $9/9$ route-matched pairs. Appendix Table~\ref{tab:api-openai} reports this result together with the controlled separation checks.}

\subsection{Ablations and Efficiency}
\label{sec:exp-ablations}

We characterize \ours's behavior as a function of the two budget parameters: the number of templates $K$ and the number of samples per template $N$. Table~\ref{tab:ablations} reports a joint $K \times N$ sweep on the $630$ checkpoint pairs of the $36$-model set.

\begin{table}[t]
  \centering
  \small
  \resizebox{\columnwidth}{!}{
  \begin{tabular}{l rrrr}
    \toprule
    & \multicolumn{4}{c}{$N$ samples per template} \\
    \cmidrule(lr){2-5}
    $K$ templates & $5$ & $10$ & $25$ & $50$ \\
    \midrule
    $5$            & $36.67\%$ & $66.67\%$ & $92.70\%$ & $97.46\%$ \\
    $10$           & $60.48\%$ & $91.90\%$ & $98.89\%$ & $99.84\%$ \\
    $20$           & $65.56\%$ & $94.13\%$ & $99.84\%$ & $100.00\%$ \\
    $30$           & $64.44\%$ & $95.24\%$ & $100.00\%$ & $100.00\%$ \\
    $40$           & $64.76\%$ & $95.71\%$ & $100.00\%$ & $100.00\%$ \\
    $60$           & $68.57\%$ & $96.03\%$ & $100.00\%$ & $100.00\%$ \\
    \bottomrule
  \end{tabular}
  }
  \caption{Template ($K$) $\times$ sample-count ($N$) ablation for \ours\ (tool-selection probe, $B=1{,}000$, $\alpha=0.05$). Cells report the rejection rate (accuracy) on the $630$ distinct-checkpoint pairs of the $36$-model set (\S\ref{sec:exp-controlled}). \CR{Accuracy generally increases along both axes, with small finite-sample fluctuations at low $N$, and saturates along a frontier: the $K=20$, $N=50$ configuration reaches $100\%$, as does $K \ge 30$ at $N \ge 25$.} At fixed small $N$, adding templates beyond $K=20$ yields only marginal gains (e.g.\ $N=5$ rises from $65.6\%$ at $K=20$ to $68.6\%$ at $K=60$); more samples per template is the more effective margin.}
  \label{tab:ablations}
\end{table}

\paragraph{Joint saturation.} \CR{The rejection rate on distinct-checkpoint pairs generally increases along both axes, with small finite-sample fluctuations at low $N$.} Holding $K$ fixed and increasing $N$ tightens the per-template categorical estimate, lifting the rejection rate until it saturates at the true separation between the two endpoints. Holding $N$ fixed and increasing $K$ adds independent template blocks to the MMD statistic, with a similar effect. The two effects compose: a small budget can be spent either as more templates with fewer samples or fewer templates with more samples. \CR{The $K = 20$, $N = 50$ configuration reaches $100\%$ on the $36$-model set and matches MET and RUT on identity discrimination.} Type-I error remains controlled across the entire grid: only $1$ of $36$ models is falsely rejected at a single cell ($K{=}10$, $N{=}10$), with all other cells at $0/36$ (Appendix Table~\ref{tab:ablations-selftest}).

\paragraph{Budget estimate.} \CR{A full audit at $K = 20$, $N = 50$ costs $1{,}000$ queries per endpoint.} A lighter $K = 10$, $N = 50$ operating point costs $500$ queries and still reaches $99.84\%$. At typical API pricing both points are well under \$$1$ per endpoint.

\subsection{\CR{Tool-Name and Schema Robustness}}
\label{sec:exp-probe-surface}

\CR{We test whether \ours depends on tool-schema wording on $12$ local and frontier models listed in Appendix~\ref{sec:ablation-details}. Holding each probe otherwise fixed, we either replace every tool name with a random neutral string or paraphrase every tool and parameter description. All three probe surfaces accept all $12/12$ self-comparisons and reject all $66/66$ distinct-model pairs (Table~\ref{tab:probe-surface-ablation}), showing that the decisions do not depend on the original tool names or wording in this experiment.}

\begin{table}[H]
\centering
\scriptsize
\setlength{\tabcolsep}{4pt}
\begin{tabular}{lcc}
\toprule
\CR{Probe surface} & \CR{Self (acc)} & \CR{Distinct (rej)} \\
\midrule
\CR{Original schema} & \CR{$12/12$} & \CR{$66/66$} \\
\CR{Random neutral tool names} & \CR{$12/12$} & \CR{$66/66$} \\
\CR{Paraphrased tool/parameter descriptions} & \CR{$12/12$} & \CR{$66/66$} \\
\bottomrule
\end{tabular}
\caption{\CR{Tool-name and schema robustness.}}
\label{tab:probe-surface-ablation}
\end{table}

\subsection{\CR{Robustness to Adaptive Providers}}
\label{sec:exp-adaptive}

\CR{We evaluate three adaptive evasion strategies: prompt-based mimicry, LoRA fine-tuning, and probabilistic substitution. We also discuss exact per-probe replay by a fully adaptive provider that recognizes audit probes.}

\noindent\CR{\textbf{Prompt-based mimicry.} We wrap each substitute in adversarial system prompts with four knowledge levels: none, claimed identity, the reference's aggregate tool-call profile, and few-shot reference demonstrations from disjoint probes. AgentProv rejects both substitutes under every condition (Table~\ref{tab:adaptive-prompt}), showing that prompt-level mimicry does not reproduce the reference tool-selection behavior.}

\begin{table}[H]
\centering
\small
\setlength{\tabcolsep}{3.5pt}
\begin{tabular*}{\columnwidth}{@{\extracolsep{\fill}}lcc@{}}
  \toprule
  \CR{Prompt knowledge} &
  \CR{\shortstack{GPT-4o-mini \\ $\rightarrow$ GPT-4o}} &
  \CR{\shortstack{GPT-5.4-mini \\ $\rightarrow$ GPT-5.4}} \\
  \midrule
  \CR{No mimicry} & \CR{rej, $6.4{\times}10^{-5}$} & \CR{rej, $5.8{\times}10^{-5}$} \\
  \CR{Identity} & \CR{rej, $6.4{\times}10^{-5}$} & \CR{rej, $5.7{\times}10^{-5}$} \\
  \CR{+Profile} & \CR{rej, $7.5{\times}10^{-5}$} & \CR{rej, $8.7{\times}10^{-5}$} \\
  \CR{+Few-shot} & \CR{rej, $8.2{\times}10^{-5}$} & \CR{rej, $1.05{\times}10^{-4}$} \\
  \bottomrule
\end{tabular*}
\caption{\CR{Prompt-based mimicry.}}
\label{tab:adaptive-prompt}
\end{table}

\noindent\CR{\textbf{Fine-tuning mimicry.} Following fingerprint-spoofing attacks using parameter-efficient fine-tuning~\citep{DBLP:journals/corr/abs-2606-16100}, we LoRA-distill Qwen2.5-1.5B on Qwen2.5-7B tool-call outputs using $48$ probe templates and test $12$ disjoint templates. The substitute evades detection on training probes but is rejected on held-out probes (Table~\ref{tab:adaptive-finetune}), so the learned mimicry is probe-specific and does not transfer to unseen probes.}

\begin{table}[H]
\centering
\small
\setlength{\tabcolsep}{10pt}
\begin{tabular}{lcc}
  \toprule
  \CR{Probe set} & \CR{Decision} & \CR{$p$-value} \\
  \midrule
  \CR{Training} & \CR{acc} & \CR{$1.00$} \\
  \CR{Held-out} & \CR{rej} & \CR{$0.013$} \\
  \bottomrule
\end{tabular}
\caption{\CR{LoRA fine-tuning mimicry.}}
\label{tab:adaptive-finetune}
\end{table}

\noindent\CR{\textbf{Probabilistic substitution.} Following randomized and rerouted substitution threats~\citep{DBLP:journals/corr/abs-2504-04715,Zhu2025RUT}, we serve the reference with probability $q$ and the substitute with probability $1-q$. The discrepancy decreases with $q$, but evasion requires serving the genuine reference for a substantial fraction of requests (Table~\ref{tab:adaptive-routing}), limiting the benefit of substitution.}

\begin{table}[H]
\centering
\small
\setlength{\tabcolsep}{3.5pt}
\begin{tabular*}{\columnwidth}{@{\extracolsep{\fill}}ccc@{}}
  \toprule
  \CR{$q$} &
  \CR{\shortstack{GPT-4o-mini \\ $\rightarrow$ GPT-4o}} &
  \CR{\shortstack{GPT-5.4-mini \\ $\rightarrow$ GPT-5.4}} \\
  \midrule
  \CR{$0.0$} & \CR{rej, $6.7{\times}10^{-5}$} & \CR{rej, $6.0{\times}10^{-5}$} \\
  \CR{$0.2$} & \CR{rej, $4.5{\times}10^{-5}$} & \CR{rej, $4.1{\times}10^{-5}$} \\
  \CR{$0.3$} & \CR{rej, $3.6{\times}10^{-5}$} & \CR{rej, $3.1{\times}10^{-5}$} \\
  \CR{$0.4$} & \CR{rej, $2.9{\times}10^{-5}$} & \CR{acc, $2.4{\times}10^{-5}$} \\
  \CR{$0.5$} & \CR{acc, $2.5{\times}10^{-5}$} & \CR{acc, $2.0{\times}10^{-5}$} \\
  \CR{$0.7$} & \CR{acc, $1.2{\times}10^{-5}$} & \CR{acc, $8.4{\times}10^{-6}$} \\
  \CR{$1.0$} & \CR{acc, $7.5{\times}10^{-6}$} & \CR{acc, $3.5{\times}10^{-6}$} \\
  \bottomrule
\end{tabular*}
\caption{\CR{Probabilistic substitution. $q$ is the fraction of requests served by the genuine reference.}}
\label{tab:adaptive-routing}
\end{table}

\noindent\CR{\textbf{Fully adaptive replay.} Exact per-probe replay is undetectable on probes to which the provider has adapted, but requires recognizing audit probes and holding the reference's per-probe distributions. We mitigate both requirements with a separately scored private probe subset. The provider cannot pre-adapt to unseen probes and therefore falls back to the substitute, which AgentProv detects. Because the probes are ordinary tool-use tasks, auditors can generate and rotate fresh private probes at negligible cost, limiting the durability of adaptation.}

\section{Conclusion}
\label{sec:conclusion}
\CR{We introduced \ours, a two-sample equality test that audits agentic LLM APIs by comparing tool-selection behavior rather than natural-language completions. The resulting categorical distribution reflects the model's post-training and learned parameters, providing a practical identity signal. Across the evaluated models, \ours\ distinguishes genuine model changes while being more robust than text-channel baselines to the deployment-context changes tested. \ours\ offers a practical, lightweight auditing method for agentic APIs, including action-only serving stacks.}

\section*{Limitations}
\label{sec:limitations}
\ours\ requires reference access to an authoritative instance of the claimed model: for fully closed models without any trusted reference channel, the test is inapplicable. \CR{As with other black-box methods, \ours\ cannot prove the underlying backbone without access to model weights. It provides strong evidence of identity from the probed tool-selection behavior, not a complete identity certificate. Within this scope, the signal is robust to the benign serving changes tested, sensitive to model substitution, and resistant to imitation on unseen probes.} Finally, the third-party API analyses cross-check \CR{the methods' decisions} against an independent token-count side-channel but do not establish ground truth on serving configuration, since providers do not disclose backbone identity or system-prompt content.

\section*{Ethical Considerations}

\CR{AgentProv is intended to support transparency and accountability
in commercial LLM deployments. A rejection at $\alpha = 0.05$
indicates a statistically significant difference in tool-selection
behavior, not proof of model substitution or provider misconduct,
because serving configuration can also affect this behavior
(\S\ref{sec:exp-hidden-prompt}). Auditors should therefore interpret
results alongside documented serving configurations, per-template
analyses (Eq.~\ref{eq:block-decomp}), and corroborating side-channels
(Appendix~\ref{sec:or-scan-method}), and should respect endpoint terms
of service and rate limits. We will release the probe templates and
source code to enable independent replication.}

\bibliography{main}

\clearpage
\appendix
\raggedbottom
\section{Technical Appendices and Supplementary Material}
\label{sec:appendix}

\subsection{Method Details}
\label{sec:method-details}

\paragraph{Action bucketing.} The auditor classifies each sample by mapping the emitted tool-call name to an option label via a per-template \texttt{action\_map}. Responses that answer directly without a tool call are bucketed into \texttt{\_no\_call}; responses whose tool-call syntax fails to parse are bucketed into \texttt{\_malformed}. Both labels are included in every option set $\mathcal{O}_k$ and are never silently discarded. Because classification reads the structured \texttt{tool\_calls[]} field rather than free-form text, bucketing is deterministic and identical across endpoints, including action-only serving stacks where \texttt{message.content} is empty.

\paragraph{Confound controls.} Three mandatory controls apply to every template. First, tool order in the schema is randomized per sample with a deterministic seed, eliminating first-listed bias. Second, tool descriptions within a template are length-matched and structurally parallel, preventing detail-driven selection. Third, prompts use neutral recipients (\texttt{@example.com}, generic names) and tool-agnostic task wording (``send a note to'' rather than ``email'') to avoid domain priming.

\subsection{Probe Examples and Template Sourcing}
\label{sec:probe-examples}

\paragraph{Template domains.} The $K$ templates span two complementary domains. \emph{General tools} cover everyday operations (weather lookup, calendar management, email dispatch, customer-info retrieval) where tool names are drawn from real provider documentation (OpenAI, Anthropic, Google) and public benchmarks, creating synthetic redundancy across $15$ task categories. \emph{MCP servers} present the model with equivalent operations from different MCP server families (e.g., \texttt{mcp\_\_calendar\_\_add\_event} vs.\ \texttt{mcp\_\_google\_calendar\_\_create\_event} vs.\ \texttt{mcp\_\_outlook\_calendar\_\_create\_meeting}), leveraging the standardized MCP protocol surface~\citep{Anthropic2024MCP}. Using tool names and parameter shapes from real SFT and RL data mixtures is essential: the probe reads the pre-trained tool-selection prior, not an arbitrary classification task. Table~\ref{tab:probe-example-templates} shows one representative template from each domain.

\begin{table*}[h]
  \centering
  \small
  \begin{tabular}{p{2.2cm}p{6.3cm}p{5.5cm}}
    \toprule
    Domain & Probe setup & Option set \\
    \midrule
    General tools &
    ``Schedule a team sync for Thursday at 2pm.'' Tools: \texttt{create\_event}, \texttt{schedule\_meeting}, \texttt{add\_event}, \texttt{book\_calendar\_event} (identical descriptions). &
    \{tool names, \texttt{\_no\_call}, \texttt{\_malformed}, \texttt{\_other}\} \\[3pt]

    MCP servers &
    ``Send a short greeting to alice@example.com.'' Tools: \texttt{mcp\_\_gmail\_\_send\_email}, \texttt{mcp\_\_outlook\_\_send\_email}, \texttt{mcp\_\_smtp\_\_send\_mail} (identical descriptions). &
    \{tool names, \texttt{\_no\_call}, \texttt{\_malformed}, \texttt{\_other}\} \\
    \bottomrule
  \end{tabular}
  \caption{Two representative templates of the tool-selection probe, showing the probe setup and the option set into which the model's response is classified. The first uses general-tool naming variants; the second uses MCP server naming variants. The full template pool contains $K = 60$ such templates spanning both domains.}
  \label{tab:probe-example-templates}
\end{table*}

\subsection{Model List}
\label{app:model-list}

\begin{table*}[h]
  \centering
  \small
  \begin{tabular}{ll}
    \toprule
    Family & Members evaluated \\
    \midrule
    Qwen$2.5$              & $0.5$B, $1.5$B, $3$B, $7$B; DistilQwen$2.5$-$3$B \\
    Qwen$3$                & $0.6$B, $1.7$B, $4$B, $4$B-Instruct-$2507$ \\
    Qwen$3$-Think          & $4$B-Thinking-$2507$ \\
    Llama-$3.1$ / $3.2$    & Llama-$3.1$-$8$B; Llama-$3.2$-$1$B, $3.2$-$3$B; SuperNova-Lite \\
    Mistral                & Mistral-$7$B-Instruct, Mistral-Nemo, Ministral-$8$B \\
    DeepSeek base          & DeepSeek-LLM-$7$B, DeepSeek-Coder-$7$B \\
    DeepSeek-R$1$ distill  & R1-Distill-Qwen-$1.5$B, $7$B, $14$B; R1-Distill-Llama-$8$B \\
    Hermes (NousResearch)  & Hermes-$3$-Llama-$3.1$-$8$B, Hermes-$3$-Llama-$3.2$-$3$B, Hermes-$2$-Pro-Mistral-$7$B \\
    Hammer-$2.1$ (MadeAgents) & $0.5$B, $1.5$B, $3$B, $7$B \\
    Phi                    & Phi-$4$-mini-reasoning \\
    Gemma-$3$              & Gemma-$3$-$1$B \\
    GLM                    & GLM-Z$1$-$9$B \\
    Other                  & Nemotron-Nano-$4$B, Gorilla-OpenFunctions-v$2$, SmolLM$3$-$3$B \\
    \CR{OpenAI (closed)}   & \CR{GPT-$3.5$-turbo, GPT-$4.1$-mini, GPT-$4$o-mini, GPT-$4$o, o$4$-mini} \\
                           & \CR{GPT-$5.4$-mini, GPT-$5.4$, GPT-$5.6$-luna, GPT-$5.6$-sol} \\
    \bottomrule
  \end{tabular}
  \caption{\CR{The $36$ open-weight checkpoints and nine closed-source OpenAI models evaluated. The five small checkpoints revisited under injected system prompts in \S\ref{sec:exp-hidden-prompt} are a subset (Llama-$3.2$-$1$B, Qwen$3$-$1.7$B, Qwen$2.5$-$1.5$B, Gemma-$3$-$1$B, Hammer$2.1$-$1.5$B). Third-party API endpoints audited in \S\ref{sec:exp-api} are listed in Table~\ref{tab:api}.}}
  \label{tab:models}
\end{table*}

\subsection{Full Pairwise Heatmap}
\label{app:full-heatmap}

\begin{figure*}[h]
  \centering
  \includegraphics[width=\textwidth]{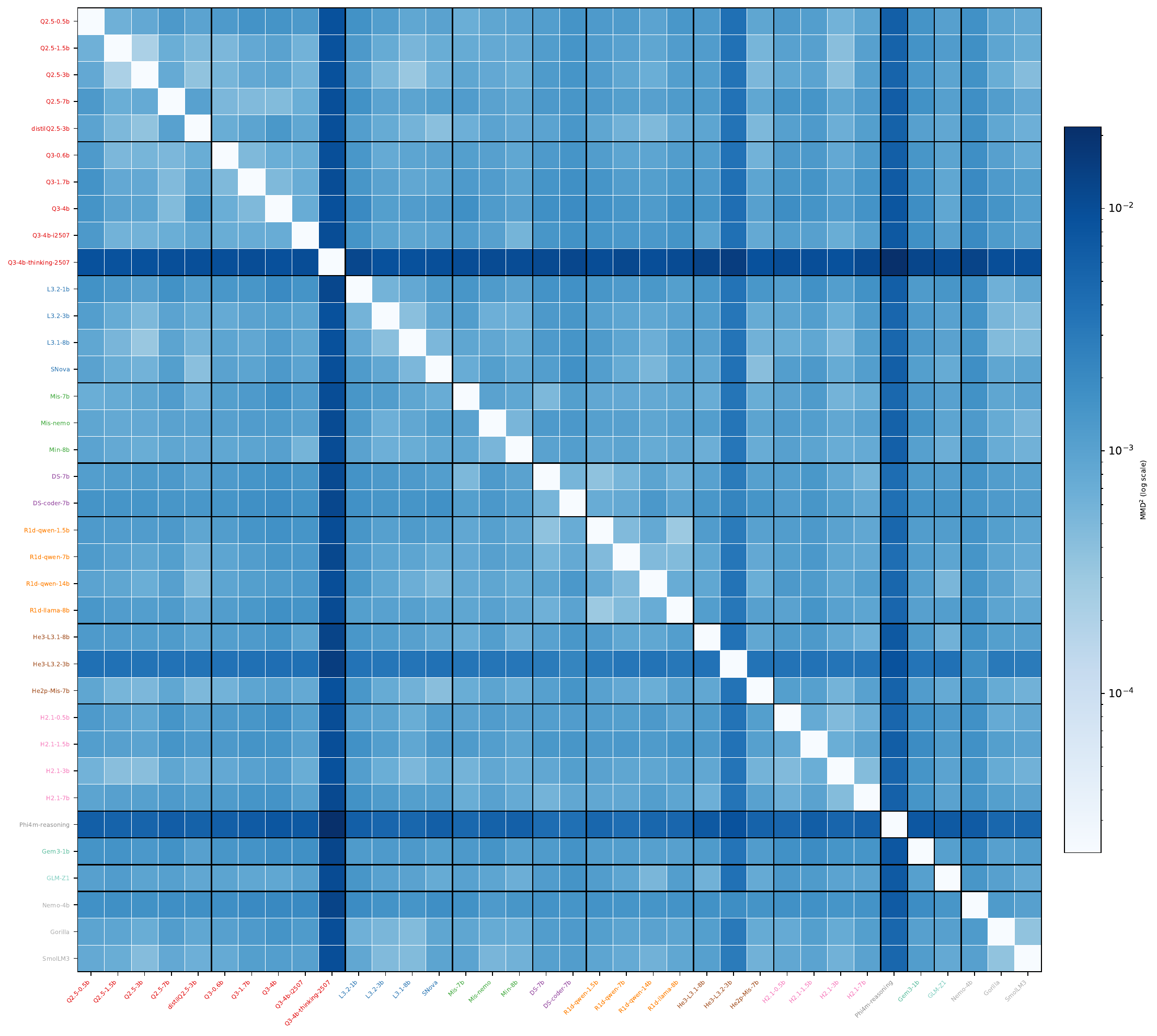}
  \caption{Full $36 \times 36$ pairwise MMD$^2$ matrix on the AgentProv $\cap$ MET $\cap$ RUT common set (\S\ref{sec:exp-controlled}), computed with \ours's tool-selection probe at $K=20$ templates and $N=50$ samples per template. Models are grouped by family; black lines mark family boundaries. All $630$ distinct-checkpoint pairs reject at $\alpha = 0.05$.}
  \label{fig:full-heatmap}
\end{figure*}

Figure~\ref{fig:full-heatmap} shows the full $36 \times 36$ pairwise MMD$^2$ matrix on the $36$-model set (\S\ref{sec:exp-controlled}), grouped by family. All $630$ off-diagonal pairs reject at $\alpha = 0.05$. Beyond the binary decision, the statistic itself reveals structure: same-family models cluster (Qwen$2.5$ sizes against each other, Llama-$3$ sizes against each other, Hammer-$2.1$ sizes against each other), visible as darker diagonal blocks. Shared training lineage also carries across architectures (R1-Distill variants distilled from DeepSeek-R1 cluster together regardless of base). Families assembled from heterogeneous bases (Hermes-$3$ SFT applied to three unrelated bases) form the exception: intra-family distances are no smaller than cross-family, reflecting that surface SFT does not override the base's tool-selection prior.

\subsection{OpenRouter Token-Count Side-Channel}
\label{sec:or-scan-method}

The token-count side-channel referenced in \S\ref{sec:exp-hidden-prompt} and \S\ref{sec:exp-api} detects provider-injected hidden system prompts by comparing the provider's reported \texttt{prompt\_tokens} against the canonical \texttt{apply\_chat\_template} count for the messages the auditor sent. A persistent positive offset indicates content the provider added that the auditor never specified.

\paragraph{Endpoint coverage.} OpenRouter exposes $358$ endpoints via \texttt{GET /v1/models}; $127$ map to a HuggingFace tokenizer release via family rules and per-endpoint overrides. After tokenizer-load and chat-template verification, $84$ endpoints remain \emph{mappable}: the auditor can independently compute their canonical chat-template token count. Of these, $75$ open-weight chat endpoints constitute the scan corpus.

\paragraph{Bare token count.} The reference is the official chat template applied to the auditor's messages, after stripping auto-injected blocks (e.g., Llama-$3.1$+ ``Cutting Knowledge Date / Today Date'') that the local template inserts but that providers may legitimately omit. A nonzero \texttt{api\_count $-$ bare\_count} therefore reflects content the provider added or removed beyond the auditor's messages, not differences in template chrome handling.

\paragraph{Probes.} For each endpoint the auditor sends four messages at temperature $0$, three calls per message: an empty user message, a single-character user message, an empty system + single-character user message, and a $20$-token continuation prompt. The headline $44\%$ figure uses the realistic prompts only (single-character user and $20$-token continuation), since edge-case prompts expose provider quirks orthogonal to deployment-context modification.

\paragraph{Classification.} For each prompt $p$, let $d_p$ denote the difference between the API-reported prompt-token count and the locally-computed bare count, and let $v_p$ denote the call-to-call variance of the API-reported count. Endpoints are classified as follows: \texttt{no\_drift} if all $d_p = 0$ and $v_p = 0$; \texttt{const\_offset\_positive} if all $d_p = k > 0$ with $v_p = 0$, indicating a consistent hidden-content injection of size $k$; \texttt{const\_offset\_negative} if $d_p = k < 0$, indicating a leaner provider template; \texttt{variable} if $v_p > 0$, indicating provider non-determinism (e.g., load-balanced deployments); and \texttt{systematic} for all other deterministic nonzero patterns.

\paragraph{Identified hidden-prompt endpoints.} Nine \texttt{const\_offset\_positive} endpoints account for the headline injections, sorted by size: Llama-$3.2$-$3$B-Instruct ($+25$), Mixtral-$8$x$22$B-Instruct ($+12$), Gemma-$3$n-e$4$b-it ($+7$), Hunyuan-a$13$B-Instruct ($+6$), and five endpoints with $+1$ token (Llama-$3$-$70$B-Instruct, Llama-$3$-$8$B-Instruct, Llama-$3.2$-$1$B-Instruct, Qwen$3$-$14$B, GLM-$4.5$v). One further endpoint (Llama-Guard-$4$-$12$B) falls into \texttt{variable} with positive mean offset, consistent with hidden content injected on a subset of calls.

\subsection{\CR{OpenAI Reseller Control}}
\label{sec:openai-reseller}

\CR{We compare the nine closed-source OpenAI models from \S\ref{sec:exp-controlled} through two routes: OpenAI's official API and OpenRouter. Using the official endpoint as the reference, \ours accepts all $9/9$ matching OpenRouter routes. This live-endpoint comparison is supporting rather than definitive evidence of faithful resale because OpenRouter's internal serving configuration remains unobservable. The controlled checks reject all $36/36$ distinct OpenAI pairs and all $324/324$ OpenAI--local-checkpoint pairs.}

\begin{table}[H]
  \centering
  \footnotesize
  \setlength{\tabcolsep}{3pt}
  \resizebox{\columnwidth}{!}{%
  \begin{tabular}{lcc}
    \toprule
    \CR{Comparison} & \CR{Pairs} & \CR{Decision} \\
    \midrule
    \CR{Matching model: official OpenAI vs. OpenRouter} & \CR{$9$} & \CR{$9/9$ acc} \\
    \CR{Distinct official OpenAI models} & \CR{$36$} & \CR{$36/36$ rej} \\
    \CR{Official OpenAI models vs. local checkpoints} & \CR{$324$} & \CR{$324/324$ rej} \\
    \bottomrule
  \end{tabular}
  }
  \caption{\CR{Closed-model reseller control and controlled separation checks. The official OpenAI endpoint is the trusted reference for the route-matched comparisons.}}
  \label{tab:api-openai}
\end{table}

\subsection{Ablation Details}
\label{sec:ablation-details}

Table~\ref{tab:ablations} in the main text reports the cross-pair rejection rate for the $K \times N$ sweep. Table~\ref{tab:ablations-selftest} complements it with the corresponding Type-I error (self-test false-positive rate) on the same grid.

\begin{table}[H]
  \centering
  \footnotesize
  \setlength{\tabcolsep}{4pt}
  \begin{tabular}{l rrrr}
    \toprule
    & \multicolumn{4}{c}{$N$ samples per template} \\
    \cmidrule(lr){2-5}
    $K$ templates & $5$ & $10$ & $25$ & $50$ \\
    \midrule
    $5$            & $0/36$ & $0/36$ & $0/36$ & $0/36$ \\
    $10$           & $0/36$ & $1/36$ & $0/36$ & $0/36$ \\
    $20$           & $0/36$ & $0/36$ & $0/36$ & $0/36$ \\
    $30$           & $0/36$ & $0/36$ & $0/36$ & $0/36$ \\
    $40$           & $0/36$ & $0/36$ & $0/36$ & $0/36$ \\
    $60$           & $0/36$ & $0/36$ & $0/36$ & $0/36$ \\
    \bottomrule
  \end{tabular}
  \caption{Self-test Type-I error (split-sample FPR) on the $36$-model set, for the same $K \times N$ grid as Table~\ref{tab:ablations}. Each cell reports the number of models for which the test rejects when both halves come from the same model (any rejection is a false positive on identity). The test is conservative: FPR is at or below the nominal $\alpha = 0.05$ rate at every cell, with the single false positive at $K{=}10$, $N{=}10$ within the expected rate by chance across the $24$-cell grid.}
  \label{tab:ablations-selftest}
\end{table}

\paragraph{\CR{Robustness model set.}} \CR{The tool-name and schema robustness experiments in \S\ref{sec:exp-probe-surface} use Qwen2.5-7B, Llama-3.1-8B, Mistral-7B, DeepSeek-7B, Hammer2.1-7B, GPT-3.5-turbo, GPT-4.1-mini, GPT-4o-mini, GPT-4o, GPT-5.4-mini, GPT-5.4, and o4-mini.}

\subsection{\CR{Additional Robustness Experiments}}
\label{sec:additional-robustness}

\paragraph{\CR{Quantization.}} \CR{We compare full-precision references with quantized deployments of Qwen2.5-7B and Llama-3.1-8B served through vLLM. Table~\ref{tab:quantization} shows that all three methods accept both $8$-bit deployments. At $4$ bits, \ours accepts Qwen2.5-7B but rejects Llama-3.1-8B, while MET and RUT reject both. The $4$-bit outcome is therefore mixed and model-dependent rather than uniformly invariant.}

\begin{table}[H]
\centering
\small
\setlength{\tabcolsep}{5pt}
\begin{tabular}{lccc}
\toprule
\CR{Quantized deployment} & \CR{\ours} & \CR{MET} & \CR{RUT} \\
\midrule
\CR{Qwen2.5-7B, 8-bit} & \CR{acc} & \CR{acc} & \CR{acc} \\
\CR{Llama-3.1-8B, 8-bit} & \CR{acc} & \CR{acc} & \CR{acc} \\
\CR{Qwen2.5-7B, 4-bit} & \CR{acc} & \CR{rej} & \CR{rej} \\
\CR{Llama-3.1-8B, 4-bit} & \CR{rej} & \CR{rej} & \CR{rej} \\
\bottomrule
\end{tabular}
\caption{\CR{Decisions for quantized deployments against their full-precision references.}}
\label{tab:quantization}
\end{table}

\paragraph{\CR{Safety policy and agent wrapper.}} \CR{We prepend two serving-level system prompts to the probes: a helpful, harmless, and honest safety preamble, and a production agent-framework protocol that requests one well-formed call from a fixed tool registry. We test Llama-3.2-1B, Qwen3-1.7B, Qwen2.5-1.5B, Hammer2.1-1.5B, Qwen2.5-7B, and Llama-3.1-8B against their clean fingerprints. Because the weights are unchanged, every rejection is a false positive. Table~\ref{tab:serving-prompts} shows that \ours has no false positives under the safety policy and two under the agent wrapper, compared with four for MET and six for RUT under either condition. The two wrapper failures show that wrapper logic can still alter the probed tool-selection behavior for some models.}

\begin{table}[H]
\centering
\small
\setlength{\tabcolsep}{5pt}
\begin{tabular}{lccc}
\toprule
\CR{System-prompt condition} & \CR{\ours} & \CR{MET} & \CR{RUT} \\
\midrule
\CR{Safety policy} & \CR{$0/6$} & \CR{$4/6$} & \CR{$6/6$} \\
\CR{Agent wrapper} & \CR{$2/6$} & \CR{$4/6$} & \CR{$6/6$} \\
\bottomrule
\end{tabular}
\caption{\CR{False positives under safety-policy and agent-wrapper system prompts.}}
\label{tab:serving-prompts}
\end{table}

\subsection{Baseline and Protocol Details}
\label{sec:protocol-details}

\paragraph{\ours\ (this work).} Each model is queried at temperature $0.7$ via its native chat template; tool schemas are shuffled per sample with a deterministic seed. \CR{We use $K = 20$ templates of the tool-selection probe and $N = 50$ samples per template, yielding $1{,}000$ samples per endpoint.} Tool-call names are bucketed via per-template \texttt{action\_map}s (\S\ref{sec:method-details}); \texttt{\_no\_call} and \texttt{\_malformed} are valid options. For each pair of endpoints we compute the observed test statistic $T$ (Eq.~\ref{eq:test-stat}) on the per-template sample matrices and calibrate the permutation null with $B = 1{,}000$ row-level shuffles (Algorithm~\ref{alg:agentprov}). Decisions are exact permutation $p$-values at level $\alpha = 0.05$.

\paragraph{MET~\citep{Gao2025MET}.} Following the upstream implementation, each model is queried on $25$ Wikipedia continuation prompts ($5$ prompts $\times\ 5$ languages: English, German, French, Russian, Spanish) at temperature $1.0$, top-$p$ $1.0$, generating $N = 50$ completions per prompt with a maximum of $50$ output tokens. Completions are padded to a fixed character length and compared with the Hamming kernel on byte-level token strings; the test statistic is the Maximum Mean Discrepancy~\citep{Gretton2012} between the reference and suspect sample sets, with a permutation null at $B = 500$ resamples. Decisions are at $\alpha = 0.05$. The MET pipeline is run unmodified through our SLURM driver; the only added code is a thin OpenAI/OpenRouter sampling hook that the upstream did not implement.

\paragraph{RUT~\citep{Zhu2025RUT}.} RUT is a grey-box test that requires log-probabilities from the reference. The protocol follows the paper: $100$ English prompts sampled from the WildChat-$1$M dataset~\citep{Zhao2024WildChat} at temperature $0.5$, top-$p$ $1.0$, maximum $30$ output tokens; for each prompt the auditor draws $1$ target completion and $m = 100$ reference completions. Every response token is scored under the reference distribution with the log-rank score (the empirical best of the five score functions evaluated in the paper). Per prompt, the randomized rank of the target's score within the reference scores is computed, and a Cram\'{e}r--von Mises goodness-of-fit test against $U(0, 1)$ decides at $\alpha = 0.05$.


\end{document}